\documentstyle[prl,twocolumn,epsf,floats,aps]{revtex}

\begin{document}
\draft

\twocolumn[\hsize\textwidth\columnwidth\hsize\csname
@twocolumnfalse\endcsname

\title{Absence of a metallic phase in random-bond Ising models in
two dimensions:\\
applications to disordered superconductors and
paired quantum Hall states}
\author{N. Read$^{1,2}$ and Andreas W. W. Ludwig$^3$}
\address{$^1$Institute for Theoretical Physics, University of California,
Santa Barbara, CA 93106-4030 \\
$^2$\cite{permaddr}Department of Physics, Yale University, P.O. Box 208120, New
Haven, CT 06520-8120 \\
$^3$Physics Department, University of California, Santa Barbara,
CA 93106-4030}
\date{July 14, 2000}
\maketitle

\begin{abstract}
When the two-dimensional random-bond Ising model is represented as
a noninteracting fermion problem, it has the same symmetries as an
ensemble of random matrices known as class D. A nonlinear sigma
model analysis of the latter in two dimensions has previously led
to the prediction of a metallic phase, in which the fermion
eigenstates at zero energy are extended. In this paper we argue
that such behavior cannot occur in the random-bond Ising model, by
showing that the Ising spin correlations in the metallic phase
violate the bound on such correlations that results from the
reality of the Ising couplings. Some types of disorder in spinless
or spin-polarized p-wave superconductors and paired fractional
quantum Hall states allow a mapping onto an Ising model with real
but correlated bonds, and hence a metallic phase is not possible
there either. It is further argued that vortex disorder, which is
generic in the fractional quantum Hall applications, destroys the
ordered or weak-pairing phase, in which nonabelian statistics is
obtained in the pure case.
\end{abstract}

 ]

\section{Introduction}

Ising models with quenched random bonds have been considered over
many years. Negative couplings produce frustration and this is the
starting point for the spin glass problem \cite{by}. A large class
of models possess a ``Nishimori line'' in their phase diagram, on
which the internal energy is analytic \cite{n}, and the
correlation functions of the Ising spins obey certain identities
\cite{n,ldgh}. In two dimensions, the Ising model can be
represented as a noninteracting fermion problem, even when the
bonds are random \cite{bp}. The problem then reduces to something
similar to a two-dimensional (2D) tight-binding Hamiltonian with
quenched disorder. Properties of the Ising model are then related
to those of the fermion system, in particular to the fermion
Green's functions corresponding to the ``Hamiltonian'', at a fixed
``energy'', namely zero (this ``energy'' is not directly related
to the energy in the sense of the Ising Hamiltonian). Then it is
of interest to understand the properties of the fermion
eigenstates near this energy, in particular whether they are
localized or extended. In this paper, we consider such problems,
and in particular argue that a recent proposal \cite{sf} that
there exists a phase of the Ising model in which the fermion
eigenstates at zero ``energy'' are extended (a ``metallic phase'')
is ruled out. We also apply the results to paired fermion systems
as in superconductors and quantum Hall states, which map onto
similar noninteracting fermion problems.

Models of noninteracting fermions can in principle be considered
using the methods of localization theory and random matrices. A
list of symmetry classes (larger than the standard list due to
Dyson) of ensembles of matrices was introduced by Altland and
Zirnbauer (AZ) \cite{az}. The work of AZ was motivated by problems
of disordered superconductors. Within the mean-field
approximation, the fermionic quasiparticles of a superconductor
are noninteracting, thus can be described using a single-particle
formulation. The latter involves a Hamiltonian which in general
contains quenched disorder, and could be a tight-binding
Hamiltonian in 2D, for example. The energy levels of this
Hamiltonian are the excitation energies of the quasiparticles.
Once again, we may ask questions about the nature of the fermion
eigenfunctions and eigenvalues. For superconductor problems, the
natural zero of energy is a special point in the spectrum (unlike
the case of a normal metal, for example) \cite{az}.

Among the symmetry classes found by AZ, one, denoted class D,
describes disordered superconductors with broken time-reversal and
spin-rotation symmetries. The symmetries are the same as those of
the fermion problem in the two-dimensional (2D) random-bond Ising
models (RBIM), and ``energy'' for the fermions of the Ising model
corresponds to excitation energy for the fermions in the
superconductor. The nonlinear sigma model for class D \cite{az},
which in effect defines this ensemble for dimensions greater than
zero, has been shown, in the 2D case, to flow under the
renormalization group to weaker values of the coupling constant
\cite{bundschuh,sf,readgr,bsz}. The coupling constant is related
to the inverse of the thermal conductivity of the superconductor,
and this flow implies that there is a phase in which there is a
nonzero density of extended fermion eigenstates at zero excitation
energy, and a superconductor described by this model would be in a
thermal metal phase. We will refer to such a phase simply as a
metallic phase. See also Refs.\ \cite{bfgm,smitha}, respectively,
for the 1D and 3D cases.

Senthil and Fisher \cite{sf} considered possibilities for the
application (via the fermion mapping) of results for class D to 2D
RBIM's. One scenario they discussed includes a metallic region in
the phase diagram, below the Nishimori line, at relatively strong
disorder and low temperature. They suggested that such a phase
would have vanishing expectation values for both the Ising spin
(``order'') and the dual ``disorder'' variables. Another scenario
was that the metallic phase should be identified with the
zero-temperature spin-glass region of a RBIM.

The preceding statements will be formulated more precisely in the
course of this paper. Here we will begin by writing the Ising
model Hamiltonian,
\begin{equation}
\beta{\cal H} = -\sum_{ ij } K_{ij}\sigma_i \sigma_j,
\label{ham}
\end{equation}
where $\beta=1/T$ is the inverse temperature, the Ising spins
$\sigma_i=\pm 1$, $i$, $j$ label sites of the lattice, and
$K_{ij}=J_{ij}/T$ is a convenient notation for the Ising couplings
(bonds). We will assume that $J_{ij}$ is zero unless $i$, $j$ are nearest
neighbors on (say) the square lattice, and that there is a $T$-independent
probability distibution for $J_{ij}$, such that the different
nearest-neighbor bonds are statistically independent and
identically-distributed. The statistical assumptions are not crucial and
could be relaxed further, but we will see that it is important that the
$J_{ij}$ are real, not complex. The partition function is then
\begin{equation}
Z=\sum_{\{\sigma_i\}}\exp(-\beta {\cal H}),
\end{equation}
where the sum is over all spin configurations $\sigma_i=\pm 1$ for all
$i$. We will avoid discussing the boundary conditions on the lattice, or
the thermodynamic limit, since we are mainly concerned with averages over
the disorder of correlations of operators at separations that can be held
fixed and far from the boundaries as the system size is taken to infinity
after the disorder average.

We now recall a trivial fact, which will be central to the later
arguments: the Ising spin correlation function for a fixed set of
bonds $J_{ij}$,
\begin{equation}
\langle \sigma_i \sigma_j\rangle \equiv
\sum_{\{\sigma_k\}}\sigma_i \sigma_j\exp(-\beta {\cal
H}) /Z,
\end{equation}
is bounded above by $1$ and below by $-1$:
\begin{equation}
|\langle \sigma_i \sigma_j\rangle| \leq 1.
\label{bound}\end{equation} The bound is attained in the
zero-temperature limit in pure or unfrustrated models, which
include the antiferromagnetic models (all $J_{ij}<0$) on a
bipartite lattice, as well as ferromagnetic (all $J_{ij}>0$)
models. The bound follows from the Boltzmann-Gibbs probabilities
$\exp(-\beta{\cal H})/Z$ being positive (and summing to 1), due to
the reality of the couplings $J_{ij}$.

In this paper, we will discuss the statistics of the correlation
functions in the order and disorder operators in a RBIM and in the
class D nonlinear sigma model. Our central result is that in the
metallic phase, the moments of either correlation function
increase as powers of distance, which for the order (Ising spin)
correlations eventually violates the upper bound, Eq.\
(\ref{bound}). This implies that the metallic phase described by
the sigma model cannot occur in a RBIM as long as the couplings
between the Ising spins are real. Our results apply to both
non-zero and zero temperature in the Ising model. We trace the
difference between the behaviors to differences in the form of the
disorder, and suggest that the metallic phase may not after all
occur in spinless or spin-polarized superconductors, or in paired
fractional quantum Hall states with disorder.

In the remainder of this paper, we present our results. In Sec.\
\ref{symm}, we show that the Kadanoff-Ceva disorder correlation
function \cite{kc} in a RBIM has moments bounded below by one, and
that its logarithm is symmetrically distributed, whenever the
bonds are symmetrically distributed, as in an Edwards-Anderson
(EA) spin-glass model. This relatively simple result will serve to
illustrate points in the later discussion. In Sec.\ \ref{met}, we
obtain our central result, that the logarithms of the squared
order and disorder correlations in the metallic phase are normally
distributed, with mean zero and variance increasing as the
logarithm of the distance, and hence the even moments of the
correlations increase as powers of distance. Several steps are
involved to set this up. An important point that arises along the
way is that the distinctions between ensembles D, B, and BD,
introduced in Ref.\ \cite{bsz}, are not important for local
properties, such as these correlations. In Sec.\ \ref{O(1)}, we
consider another model, the O(1) model, and show that both its
order and disorder correlations have properties like those in
Sec.\ \ref{symm}. This model is most likely in the metallic phase.
The crucial difference between such a model, and the RBIM, is that
(in network model \cite{cc} language, discussed in Sec.\
\ref{met}) the disorder adds $\pi$ fluxes or vortices on one
sublattice in the RBIM, but on both in the O(1) model; in Ising
model language, the O(1) model corresponds to an Ising model with
some couplings being complex. We also obtain the exact exponent
for the mean order and disorder correlations at the critical point
in another network model, the class C, or spin quantum Hall, model
of Ref.\ \cite{kag}. In Sec.\ \ref{app}, we consider applications
of our results to spinless or spin-polarized p-wave
superconductors or paired fractional quantum Hall effect (FQHE)
states. We show that independent insertion of vortices on a single
sublattice corresponds to the RBIM situation, and cannot produce a
metallic phase, at least at low densities. We argue that such
``vortex disorder'' destroys the Ising low-temperature ordered, or
weak-pairing phase. For correlated vortices, the latter phase can
occur, and there may be transitions in the universality classes
found in the RBIM, rather than an intermediate metallic phase.
Sec.\ \ref{con} is the conclusion.

\section{Disorder Correlations for a Symmetric Distribution of
Bonds} \label{symm}

Our first result concerns the dual correlations in the EA spin
glass case where the mean of $J_{ij}$ is zero. The two-point
correlation of the Kadanoff-Ceva disorder variable $\mu_\alpha$ is
defined in the following way (adapted from the pure case
\cite{kc}). The disorder variables are associated with sites
$\alpha$ of the (graph-theoretic) dual lattice, that is plaquettes
of the original lattice. Given a choice of two such sites
$\alpha$, $\beta$, we take the Hamiltonian (\ref{ham}) and modify
it by reversing the sign of the $J_{ij}$'s on the links of the
lattice crossed by a path on the dual lattice that runs from
$\alpha$ to $\beta$. We can then construct the corresponding
modified partition function $Z_{\rm mod}$. Then we define
\begin{equation}
\langle \mu_\alpha \mu_\beta\rangle \equiv Z_{\rm mod}/Z.
\end{equation}
This definition is independent of the choice of path from $\alpha$ to
$\beta$, because of ${\bf Z}_2$-gauge properties of the Ising model. Note
that $\langle \mu_\alpha \mu_\beta\rangle>0$ when the $J_{ij}$'s are real.

Now we consider the statistical properties of the disorder correlation
function. We denote the average over the random bonds by an overbar, for
example $\overline{\langle \mu_\alpha \mu_\beta\rangle}$. We again
make use of ${\bf Z}_2$ gauge properties, this time of the distribution
function for $J_{ij}$. There is a statistical ${\bf Z}_2$ gauge invariance
if the distribution is symmetric, $P(J_{ij})=P(-J_{ij})$ for each $i$,
$j$. However, such reversed bonds were exactly what was used in the
definition of the disorder correlation. The set of bonds used in $Z_{\rm
mod}$ occurs with the same probability, or probability density, as those
in $Z$. Also, interchanging the original with the modified bonds exchanges
$Z_{\rm mod}$ with $Z$. Hence $\ln \langle \mu_\alpha \mu_\beta \rangle$ is
symmetrically distributed, and
\begin{equation}
\overline{(\ln \langle \mu_\alpha \mu_\beta \rangle)^m}=0
\end{equation}
for $m$ odd, while
\begin{equation}
\overline{(\ln \langle \mu_\alpha \mu_\beta \rangle)^m}\geq 0,
\end{equation}
for $m$ even. For the correlation function itself, we have
\begin{equation}
\overline{\langle \mu_\alpha \mu_\beta\rangle}=\overline{Z_{\rm mod}/Z}=
\frac{1}{2}(\overline{Z_{\rm mod}/Z}+\overline{Z/Z_{\rm mod}})\geq 1.
\end{equation}
The same argument works for any moment of the correlation function,
\begin{equation}
\overline{\langle \mu_\alpha \mu_\beta\rangle^m}\geq 1,
\label{mombound}
\end{equation}
for any positive or negative integer $m$. The bounds are attained in the
high temperature limit, where $\langle \mu_\alpha\mu_\beta\rangle=1$.

We can predict how the disorder correlation function would behave
in some well-known phases. In the paramagnetic phase, where
$\langle \sigma_i\sigma_j\rangle\to 0$ as $r_{ij}$ (the distance
between $i$ and $j$) goes to infinity, we expect that the mean
disorder correlation goes to a constant at large distances, as in
the pure case, and as in the high temperature limit. The constant
must be $\geq 1$, and it appears that it will increase with
decreasing temperature. We also expect that the width of the
distribution of the logarithm of the correlation goes to a
constant. A finite-temperature spin-glass phase is believed not to
occur in 2D, but if it did we would predict that the distribution
of $\ln\langle\mu_\alpha\mu_\beta\rangle$ would have a width that
goes as  $(C_1 r_{\alpha\beta}^{\theta}+ C_2)/T$ at low
temperature, where $C_1$, $C_2$ are positive constants, and
$\theta$ is an exponent that characterizes the spin-glass phase
\cite{fh} as follows. In the spin glass, the insertion of the
disorder variables induces a domain wall terminating at $\alpha$
and $\beta$. The wall is a fractal object, with a fractal
dimension less than $2$, and its free energy, which is random and
can be positive or negative, scales as $r_{\alpha\beta}^\theta$
\cite{fh}. This exponent is believed to be the same one that
enters the effect of reversing the boundary conditions, from
periodic to antiperiodic, in one direction in a finite system of
size $L$; the change in free energy scales as $L^\theta$. The
exponent $\theta$ must be positive if the spin-glass phase is to
be stable at finite $T$; it is found numerically to be negative,
for continuous (e.g. Gaussian) distribution of $J_{ij}$,
indicating that no finite $T$ spin glass phase exists in 2D
\cite{by}. For some special discrete distributions, such as the
bimodal $\pm J$ distribution (which has many degenerate ground
states, giving an extensive entropy at $T=0$), $\theta$ is small
and negative, or possibly zero \cite{kr}. Finally, for a critical
point, $r_{ij}^\theta$ in the width should be replaced by $\ln
r_{ij}$ to a power $\geq 1/2$, but $<1$, when certain conditions
hold, or most generally a function of $r_{ij}$ that is smaller
than $\ln r_{ij}$ as $r_{ij}\to \infty$ (these follow from general
results in Ref.\ \cite{ludwig90}).

The result Eq.\ (\ref{mombound}) is in stark contrast to the Ising (order)
correlations. With a symmetric probability distribution, the odd moments
of $\langle \sigma_i\sigma_j\rangle$ vanish, and the even moments are
$\leq$ 1. These opposite inequalities illustrate the extreme {\em lack} of
duality for a symmetric probability distribution. If a metallic phase did
occur in the RBIM, it would have to be due to frustration, as recognized
in Ref.\ \cite{sf}. It would then naively be expected to occur for a
symmetric distribution of bonds. We have now shown that the idea of a
phase in which the mean disorder correlation tends to zero is untenable in
any RBIM with a symmetric distribution of bonds.

What has happened to the duality present in the pure 2D Ising model?
Kramers and Wannier showed that the Ising model on the square
lattice can be reformulated as a dual model on the dual lattice, with
Ising spins $\mu_\alpha=\pm 1$, and dual couplings $\tilde{K}$. If the
disorder variables become Ising spins, why does one not again obtain a
correlation less than one? In the pure case, of course, one does. But the
general relation between the original couplings $K_{ij}$ and the
corresponding $\tilde{K}_{\alpha\beta}$ is
\begin{equation}
\exp(-2\tilde{K}_{\alpha\beta}) = \tanh K_{ij}.
\end{equation}
For $K_{ij}<0$, $\tilde{K}_{\alpha\beta}$ has an imaginary part
$i\pi/2$ (modulo a multiple of $i\pi$). The Boltzmann-Gibbs weights of the
dual spin configurations become complex in general. However, the weights
for a given nearest-neighbor bond $\alpha$, $\beta$ for the two values of
$\mu_\alpha\mu_\beta=\pm 1$ differ simply by a sign. The disorder
correlation for fixed $J_{ij}$'s then becomes a weighted average of
$\mu_\alpha\mu_\beta$ (for arbitrary $\alpha$, $\beta$) with weights that
sum to one but can be positive or negative. Hence the disorder correlation
can be larger than one. Put another way, $Z$ may be smaller than $Z_{\rm
mod}$, unlike the pure case.

For more general distributions, including those with a nonzero mean for
$J_{ij}$ (which we can take to be positive without loss of generality), we
cannot obtain a general result so easily. It is clear that when the
disorder is weak (say, the standard deviation is small compared with the
mean), there will be a ferromagnetically-ordered Ising phase, as in the pure
Ising model, and in this the disorder correlation goes to zero at large
distances. In order to rule out the existence of a metallic phase in the
intermediate region with non-zero mean $J_{ij}$, another approach is
needed.

\section{Spin Correlations in the Metallic Phase}
\label{met}

Now we turn to our second result, which directly concerns the
metallic phase in the nonlinear sigma model for class D. We ask
the question: if such a phase occurs in a random Majorana fermion
model, what will be the behavior of the order and disorder
correlations? We note immediately that the phase, as discussed in
Refs.\ \cite{sf,readgr}, is intermediate between two localized
phases that would be identified with the paramagnetic and
ferromagnetic Ising phases, which are still approximately dual to
each other as in the pure Ising model. Then the intermediate
metallic phase maps to itself under duality, and should treat the
order and disorder correlations on an equal footing. The
asymptotics of the two correlation functions should be similar.

In Sec.\ \ref{ferm}, we discuss the representation of the Ising
model as a lattice free fermion quantum field theory, the relation
of this to a network model, and the representation of order and
disorder correlations in this language. In Sec.\ \ref{nlsm}, we
describe the nonlinear sigma model that is used to define the
metallic phase. We argue that the distinctions between ensembles
D, B, and BD \cite{bsz}, that differ globally, are not important
for local correlations. In Sec.\ \ref{twist}, we introduce the
``twist operators'' that represent the order and disorder
operators in the nonlinear sigma model. Then in Sec.\
\ref{result}, we show that the statistics of the order
correlations is incompatible with a RBIM with real bonds, but
compatible with other models that violate the latter condition.

\subsection{Fermion Representation}
\label{ferm}

\begin{figure}
\epsfxsize=4in \vspace{-1.4in}
\centerline{\epsffile{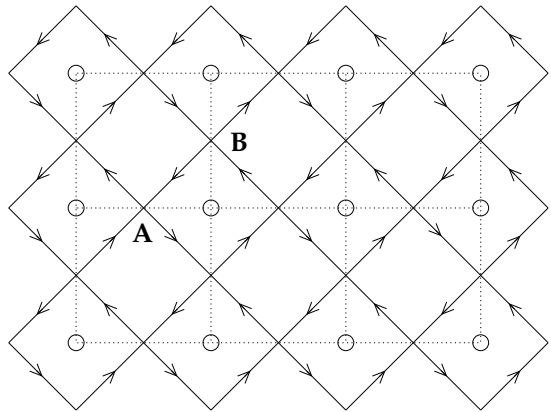}} \vspace{-1.3in}
\caption{Relation of the Ising model and the network model. Ising
spins are located at the open circles, and bonds are shown dotted.
Solid lines with arrows form the ``medial graph'', on which the
network model is defined. Examples of nodes on each of the two
sublattices, corresponding to the horizontal and vertical bonds,
are labeled $A$, $B$, respectively. Note the form of the edge of
the cluster.} \label{fig1}\vspace{-0.1in}
\end{figure}

The metallic phase in the nonlinear sigma model for class D
describes Majorana fermions, so we must consider the fermion
representation of the Ising model. This can be set up in a variety
of ways. The details are not in fact all that important here. The
important points are that in fermion language, duality becomes
rather self-evident, and both the order and disorder variables are
represented as modifying the partition function by inserting an
additional ${\bf Z}_2$ fluxes or vortices seen by the fermions. A
fermion propagating around a vortex picks up a phase factor $-1$.
The difference between the two operators is in the locations on
the lattice at which they occur. The duality is most evident if
the fermions are considered as moving on the ``medial graph'' of
the original square lattice \cite{bax}, as shown in Fig.\
\ref{fig1} for a simply connected cluster. The medial graph of a
given planar graph possesses two sublattices of plaquettes, on one
of which each plaquette encloses a site of the original lattice,
and on the other of which each plaquette encloses the center of a
plaquette of the original lattice (i.e.\ a site of the dual
lattice). The medial graph of the square lattice is another square
lattice, as shown, and we consider only square-lattice clusters
from here on. Note that we view the corners outside the cluster as
nodes, so that the total number of links in the medial graph is a
multiple of four.

The square of the Ising model partition function $Z$ can be
represented as a six-vertex model on the medial graph, with
free-fermion values of the parameters at each node \cite{bax}.
(There are also many other ways to represent the Ising model as a
noninteracting fermion field theory on a decorated version of the
square lattice. One such approach was used in pioneering work by
Blackman and Poulter \cite{bp} on the RBIM.) This free-fermion
system is also equivalent to a (second-quantized representation
of) the Chalker-Coddington network model \cite{cc}, as has been
emphasized recently \cite{chothes,grl}. In other words, the single
particle model underlying the fermion field theory is a network
model. We omit a complete description of these models since they
have been discussed so frequently in recent years, but an outline
of the main points is as follows. The links of the medial graph
square lattice are viewed as directed with an arrow on each link;
the arrows circulate around the plaquettes, which implies that
they circulate in opposite ways for the two sublattices of
plaquettes (see Fig.\ \ref{fig1}). The particle propagates on the
links of this medial graph in the direction of the arrows, picking
up amplitudes that depend on the original random bonds $J_{ij}$.
The amplitudes for each time step, during which the particle must
move ``forward'', following the arrows, to an adjacent link
consistent with the arrows on the network, are elements of a
unitary ($S$-) matrix assigned to each node. Thus the
time-evolution is described by a unitary matrix $\cal U$, that is
real in the present case, and has size a multiple of $4$. The sign
of the product of amplitudes picked up by the particle propagating
once around a plaquette determines whether a ${\bf Z}_2$ flux or
vortex (we use these terms, or $\pi$ flux, interchangeably) is
present; a vortex is present when the sign is $-1$. In the pure
Ising model, such a vortex (a flux of $\pi$) is present on every
plaquette. The insertion of negative $J_{ij}$ in the Ising model
introduces an additional pair of ${\bf Z}_2$ fluxes on the
plaquettes of the medial graph that enclose the plaquette centers
of the original lattice of the two plaquettes that are adjacent to
the bond in question \cite{bp}. When we speak of adding vortices
or fluxes to plaquettes, the fluxes add mod $2\pi$, since the net
phase picked up by the particle is what really counts; the gauge
choices involved will not matter. The effect of negative
$J_{ij}$'s in the Ising model is thus to add vortices, but {\em
only on one of the two sublattices of plaquettes} of the medial
graph network. By duality, vortices can also be produced in
similar pairs on the other sublattice, by adding an imaginary term
$i\pi/2$ to $K_{ij}$ (see Sec.\ \ref{symm}).

The squared partition function of the Ising model, $Z^2$, is now
given by the second-quantized version of the network \cite{grl}.
The partition function for noninteracting fermions is generally a
determinant of the inverse fermion propagator; in the present
case, the propagator between two links is a sum over paths, given
by the corresponding matrix element of $1+{\cal U}+{\cal
U}^2+\ldots=(1-{\cal U})^{-1}$, so we have $Z^2\propto
\det(1-{\cal U})$. Note that, unlike many other representations of
the Ising model as a fermion field theory, in our case the matrix
$1-{\cal U}$ is not antisymmetric, so we cannot say that $Z$ is
the Pfaffian of the same matrix.

Because $\cal U$ is unitary, its eigenvalues lie on the unit
circle and may be written $e^{-i\epsilon}$, where the eigenvalues
$\epsilon$ of $i\ln {\cal U}$ play the role of excitation energy
eigenvalues, even though they are defined only mod $2\pi$. It is
clear that for the long-time properties, such as the partition
function, the important part of the spectrum of $\epsilon$ is near
$\epsilon=0$. Since $\cal U$ is real, its complex eigenvalues come
in complex conjugate pairs, while $1$ and $-1$ are possible and
will usually be nondegenerate. Also, since the network has a
two-sublattice property (the particles hop from one type of link
to the other alternately), the eigenvalues come in pairs
$e^{-i\epsilon}$, $-e^{-i\epsilon}$. This implies that if $1$,
$-1$ are present, then so are $i$, $-i$, since the total number of
eigenvalues is a multiple of $4$. Thus we could restrict attention
to the range $-\pi\leq \epsilon \leq \pi$, which represents
$-\infty$ to $\infty$ in a continuum model. For the RBIM, the pair
$1$, $-1$ does not occur, $\det(1-{\cal U})>0$, and the square
root can be taken to obtain $Z>0$ \cite{bp}. The case without the
quadruplet $1$, $-1$, $i$, $-i$ (i.e. when $\det{\cal U}=1$)
corresponds to random matrices in class D, while the case with
that quadruplet, $\det{\cal U}=-1$, corresponds to those in what
has been termed class B \cite{bsz}. These random matrix ensembles
are of matrices in the Lie algebras of SO($2N$), SO($2N+1$) (for
some $N$), respectively \cite{az,bsz}. Matrices found in class B
possess at least one, and typically only one, exact zero
eigenvalue.

We now consider the calculation of the moments of the two-point
functions of the order and disorder variables in the metallic
phase of the nonlinear sigma model for class D in 2D. In terms of
the medial graph or network model, the order and disorder
operators are both represented as the ratio of a modified to the
unmodified partition function, where order variables are
represented by modifying the partition function by inserting
vortices on the sublattice of plaquettes that correspond to the
sites of the original lattice, and the disorder variables are
vortices on the plaquettes that correspond to the plaquettes of
the original lattice. Either partition function, when squared, is
given by $\det(1-{\cal U})$. As a check on the formulas, we can
consider the order and disorder correlations in the pure case. An
isolated vortex on a site of the original lattice carries a zero
eigenvalue $\epsilon=0$ in the high, but not in the low,
temperature phase. For two vortices, the zero modes can mix and
split away from zero, by an amount exponential in the separation
when the latter is greater than the correlation length. At such
large distances, the other eigenvalues tend to nonzero constants,
so the behavior of the ratio of products of eigenvalues of
$1-{\cal U}$ is determined by the eigenvalues that tend to zero.
Hence the correlation function tends to zero exponentially with
distance in the high $T$ phase, but goes to a constant in the low
$T$ phase. For the disorder correlation, the situation is
reversed.

The average over the disorder of the ratio of determinants is
performed by using either the replica method, with $2n$ copies of
the system and $n\to0$, or the supersymmetry method, where $2n$
copies of the system are supplemented by $2n$ copies of the system
with a certain kind of boson in place of the fermions, and no
$n\to0$ limit. In the supersymmetry method, the bosons cancel the
fermion determinants, as long as they are all unmodified. We will
use the replica method, but the same results can easily be
obtained with supersymmetry. For technical reasons, it is easiest
to consider only the moments with $m=$ even of the correlation
functions. Then we need to average the ratio of the $m$th power of
the modified partition function to the unmodified partition
function. Therefore we will modify the network for $m$ copies of
the fermions so that they pick up an additional factor $-1$ on
propagating around either vortex (we can take these on the
positions of the original sites, so as to obtain the spin-spin
correlation function, but the disorder correlation is similar).
The remaining $2n-m$ fermions are unmodified. When $n\to0$, the
partition function of the latter yields the division by $Z^m$.
Thus in the average, the moment of the correlation function is
simply the partition function of the replicated system at $n=0$,
that is $=1$ when unmodified, but is not when $m$ of the Majorana
fermions have been modified. That is
\begin{equation}
\overline{(\langle\sigma_i\sigma_j\rangle)^m}=\overline{(Z_{\rm
mod}/Z)^m}=\lim_{n\to0} {\cal Z}_{(m)}/{\cal Z},
\end{equation}
where $\cal Z$ stands for the partition function of the replicated and
averaged system, $\lim_{n\to 0}{\cal Z}=1$, and the subscript indicates
that $m$ components have been modified.

As we will discuss further below, the nonlinear sigma model for
the metallic phase in class D requires us to introduce an infrared
regulator $\eta>0$ which can be viewed as an imaginary part of the
energy (the real part being $=0$) at which we calculate the
fermion Green's functions, or as a corresponding shift in the
energy eigenvalues. This is necessary in random fermion problems
when the mean density of states at $\epsilon=0$ is nonzero, so in
general it can be included as a precaution. In the network model,
it can be included by replacing $\cal U$ by ${\cal U}e^{-\eta}$.
We will need, first, to take moments in a finite size system with
$\eta>0$, then take the system size to infinity, then let
$\eta\to0$. Some preliminary investigation suggests that for the
moments of the ratio of determinants we consider, with finite
separation of $i$ and $j$ (or $\alpha$ and $\beta$), this will
give the same result as taking $\eta=0$ from the beginning, which
is the strict definition for Ising models. This is for a fixed
nonzero Ising temperature $T$. However, if one tries to take the
temperature to zero before $\eta\to 0$, problems may arise. The
reason is that, in the $T\to 0$ limit, the fermions circulate
around the plaquettes of the original Ising lattice, with
amplitudes 1 (for $\eta=0$). The eigenvalues of ${\cal U}^4$ are
then determined by the flux, either $0$ or $\pi$, on those
plaquettes. Hence the eigenvalues $\epsilon$ tend to either
$4\epsilon=0$ (mod $2\pi$) or $4\epsilon= \pi$ (mod $2\pi$), and
when the RBIM has a finite probability for any given plaquette to
be frustrated, a finite fraction of eigenvalues $\epsilon$ (and
also the corresponding eigenvalues of $1-{\cal U}$) tend to zero
as $T\to 0$ \cite{bp}. In the modified partition function needed
to obtain the spin correlation squared, the number of eigenvalues
$\epsilon$ that tend to zero is the same as in the unmodified
partition function, since otherwise the spin correlation will go
to zero or infinity, which is not the case. It is only these
eigenvalues that are important in determining the spin correlation
in the $T \to 0$ limit. When the partition function is regulated
with $\eta$, the corresponding eigenvalues of $1-{\cal
U}e^{-\eta}$ tend to $\eta$, independent of $\epsilon$, and the
squared spin correlation goes to 1. This is expected in the case
of a continuous distribution of random bonds, but definitely not
for a bimodal ($\pm J$) distribution, where the $T=0$ spin
correlation should be nontrivial. Thus the order of limits $T\to
0$, $\eta \to 0$, makes a difference in this case.

\subsection{Nonlinear Sigma Model}
\label{nlsm}

The claim about the metallic phase is that, in that phase, the
partition function $\cal Z$, and correlation functions, can be
represented at large distances by those of the nonlinear sigma
model for class D \cite{bundschuh,sf,readgr}. In replica language,
this model contains a field that takes values in the target
manifold O($2n$)/U($n$). This may be parametrized by a $2n\times
2n$ complex matrix $Q$, which obeys $Q=Q^\dagger$, $Q^2=I_{2n}$,
and $Q^t=-\Lambda_x Q \Lambda_x$, where $^t$ denotes transpose,
and $\Lambda_x=I_n\otimes \tau_x$ ($\tau_x$ is a $2\times 2$ Pauli
matrix). In terms of $n\times n$ blocks, the top right block $V$
of $Q$ is an $n\times n$ antisymmetric complex matrix. (A
different parametrization is used in Ref.\ \cite{sf}.) The
symmetry operations are $Q\to O Q O^\dagger$, where in our basis,
a matrix $O$ is in O($2n$) if $O^{-1}=O^\dagger=\Lambda_x O^t
\Lambda_x$ [and in SO($2n$) if also $\det O=1$]. $Q$ can be
written as $Q=U \Lambda_z U^{-1}$ for $U$ in O($2n$), where
$\Lambda_z=I_n\otimes\tau_z$ ($U$ should not be confused with
$\cal U$). This represents the coset space O($2n$)/U($n$) because
$Q$ is invariant when $U\to U g$, where $g$ is in the U($n$)
subgroup of SO($2n$) parametrized in our basis as $g={\rm
diag}(u,u^*)$, where $u$ is a $n \times n$ unitary matrix [thus,
in U($n$)], and $u^*$ is the complex conjugate of $u$.

In Ref.\ \cite{bsz}, it was emphasized that O($2n$)/U($n$) has two
disconnected components, corresponding to whether $\det U=\pm 1$.
For a zero-dimensional system, the other ensembles, termed classes
B and BD, can be obtained by treating the component with $\det
U=-1$ differently \cite{bsz}. These correspond to the existence of
a single exact zero mode, $\epsilon=0$, in a finite size system,
with probability 1 (for class B) or $1/2$ (for class BD). Some
network models (still with real $\cal U$) possess such zero modes,
namely whenever $\det {\cal U}=-1$, and this can occur, depending
on what fluxes are present, and the boundary conditions. We can
avoid them by making appropriate choices of the latter. Even when
present, they cancel in the regularized ($\eta\neq0$) ratios of
determinants we consider in this paper. The reason is that when we
insert two additional $\pi$ fluxes on the same sublattice, the
determinant of $\cal U$ does not change. (However, an exact zero
mode could still affect the other eigenvalues through level
repulsion effects, for example.) While the presence or absence of
such a zero mode may be important in random matrix ensembles for
zero dimensional systems, or for global properties in higher
dimensions, we do not expect it to play a role in {\em local}
properties in more than zero dimensions, such as the correlations
we consider here. (This applies to localized, as well as metallic,
phases.) Therefore we expect that the distinctions between the
nonlinear sigma models should not be important, and we will refer
to class D/B/BD when this is so.

For dimensions larger than zero, a precise prescription for
handling the two components of the target manifold has not been
given. One would expect there could be domains of the two
``phases'' (in which the field $Q$ is on one or other of the two
components of the target manifold). The domain walls would likely
cost some action per unit length, and therefore additional
parameters will be needed to specify the model. We would expect
that there will then be a regime of parameters in which domain
walls are costly and all domains of the ``opposite'' phase are
small. Then the $Q$ field would essentially be globally on one
component or the other. In that case, calculations can be done
without domain walls as in other sigma models, but with a sum over
the two phases. In fact, all existing proposals for a metallic
phase in class D/B/BD \cite{bundschuh,sf,readgr} neglect domain
walls. Alternative phases where domain walls proliferate may
exist, but have not been identified, and may not be metallic. In
the absence of such proposals, we will consider the system without
domain walls as defining the metallic phase we consider here. We
note that the results in Ref.\ \cite{grl} give a way to handle, in
effect, the different components of the target manifold in a
strong-coupling situation in dimensions $\leq 2$.

We further argue that the regulator $\eta$ which we introduce
suppresses the second component. In the nonlinear sigma model, it
introduces a term in the action of the form $-\eta\int d^2r\,{\rm
tr}_{2n} \Lambda_z Q$, where ${\rm tr}_{2n}$ denotes a trace over
$2n$-dimensional space. This term has to be minimized on each
component to find the saddle point(s) about which perturbative
fluctuations are expanded. We find that at such $Q$ values for the
two components, where $Q=\Lambda_z$, $Q=O\Lambda_z O^{-1}$,
respectively, and $O$ represents a reflection in a hyperplane, the
second component has relative weight like $e^{-\eta L^2}$ compared
with the first, where $L^2$ is the area of the system. Since we
take $L^2\to\infty$ before $\eta\to 0$, we find that the second
component is suppressed. This does not change the partition
function ${\cal Z}=1$ at $n=0$ for $\eta=0$, since for $\eta=0$
the functional integral over the first component gives $1$, and
that over the second component gives $0$. (The use of just the
first component, which includes $U=I_{2n}$, corresponds strictly
to class BD \cite{bsz}.) Therefore, we drop the second component
entirely, and no difference between the metallic phases in classes
D, B, and BD will be seen in local correlations. (In the total
density of states in the ``ergodic'' regime discussed in Ref.\
\cite{bsz}, smearing by energy resolution $\eta$ makes all three
classes the same when $\eta$ is greater than the level spacing, of
order $L^{-2}$, consistent with this conclusion.)

\subsection{Twist Operators}
\label{twist}

{}From a perturbative point of view, $Q$ arises from bilinears in
the underlying Majorana fermions, which naturally leads to
antisymmetric matrices. To obtain the correct structure of $Q$, it
is essential that we start from the correct vacuum at weak
coupling, represented by $Q=\Lambda_z$, which is invariant under
the U($n$) subgroup introduced above. The basis in which we gave
$Q$ corresponds to the use of $n$ complex Fermi fields $\psi$ in
place of the $2n$ Majoranas $\xi$. The diffusing (Goldstone) modes
of the model involve only modes of the form $\psi\psi$ or
$\psi^\dagger\psi^\dagger$ (the indices are suppressed), which
correspond to the two off-diagonal blocks. (Goldstone modes
corresponding to $2n\times 2n$ real antisymmetric matrices would
give class DIII \cite{az}, in which time-reversal is unbroken.)
This parametrization can also be arrived at using the O($1$)
network model, which in first-quantized form is a single particle
propagating on the medial graph network with fixed nodes of a
standard form \cite{cc}, and picking up $\pm 1$ factors (with
independent probabilities $1/2$) on each link. Averaging over the
group O(1)$\cong {\bf Z}_2$ in a replicated second-quantized
representation leads to propagating Goldstone modes, and this
model is in class BD, as described in Refs.\ \cite{bsz,ohone}.

In general, the modified partition function of the model is defined by the
presence of a ``twist'' in the $Q$ field. The twist is a boundary
condition at the points corresponding to $i$, $j$, that is obtained from
the fact that $m$ of the Majorana fermion fields pick up a $-1$. Since
$m$ is even, this corresponds to a proper rotation $O$ in SO($2n$).
We can choose the $m$ components of the fermions that are modified to be
the real and imaginary parts of the first $m/2$ of the complex fermions
that define our basis for $Q$. Then $O$ is represented by a matrix in the
same U($n$) subgroup mentioned above, with $O={\rm diag}(u,u^*)$, and
$u={\rm diag}(-1,-1,\ldots,1,1\ldots)$ with $-1$ appearing $m/2$ times.
Hence the modified partition function ${\cal Z}_{(m)}$ is defined as the
usual one but with the condition on the $Q$ fields at the points $i$, $j$,
that on making a circuit around these points the $Q$ field is not periodic
but changes as $Q\to O Q O^\dagger$, using the same $O$.

\subsection{Result at Weak Coupling}
\label{result}

As mentioned above, the nonlinear sigma model for class D/B/BD
flows (when $n=0$) to weak coupling. Accordingly, we can compute
the spin correlation function in the weak-coupling limit. To
leading order, the action can be approximated as Gaussian for
small $V$,
\begin{equation}
S=\frac{1}{2g^2}\int d^2r\,{\rm tr}_n\, \nabla V \nabla V^\dagger,
\end{equation} where the trace is over the $n\times n$ matrices $V$, and
$g^2$ here is the coupling constant squared, proportional to the
inverse of the thermal conductivity $\kappa_{xx}$
\cite{sf,readgr}. We have neglected the topological ($\theta$)
term, since it plays no role in the following calculation. We have
also omitted the leading nontrivial part of the $\eta$ term,
$\eta\int d^2r\,{\rm tr}_n \,V V^\dagger$ with $\eta>0$. The limit
$\eta\to 0$ is taken after the thermodynamic limit, because
massless scalar fields in an infinite 2D system are problematic.
In the weak-coupling limit, the twist operators take a simple
form, since the operation described by our choice of $O$ acts
linearly on $V$; $V$ transforms as the antisymmetric second-rank
tensor representation of U($n$). The condition on $V$ on going
around the points $i$, $j$ is that the components corresponding to
complex fermions that are both modified or both unmodified are
periodic, but those corresponding to one modified and one
unmodified fermion pick up $-1$. Thus $m(2n-m)/4$ distinct
(complex) components of $V$ pick up a $-1$ on going around $i$ or
$j$, and the remainder of the total $n(n-1)/2$ pick up $+1$ (are
periodic).

We now need the ratio of the modified to unmodified partition
functions for the $V$ field with $n\to0$. This has the form of a
standard problem in conformal field theory \cite{dfms} (the
Gaussian theory is conformal since the coupling $g$ does not get
renormalized). A twist operator of a single real massless scalar
field, defined as a ratio of partition functions as here, has
conformal weight $1/16$, and so its left-right symmetric
correlation function decays as $r^{-1/4}$. The exponent is doubled
for a complex scalar, both of whose components are twisted.
Multiplying these for our $m(2n-m)/4$ complex components, we
obtain
\begin{equation}
\overline{(\langle \sigma_i\sigma_j\rangle)^m}\sim
r_{ij}^{-m(2n-m)/8}, \label{mom} \end{equation} which is the
central result of this paper. Note that this result is independent
of the coupling $g$. When $n \to 0$, we obtain $r^{m^2/8}$, a {\em
positive} power of distance. In the full non-linear sigma model,
$g^2$ approaches zero logarithmically with distance when $n=0$
\cite{bundschuh,sf,readgr}. {}From standard perturbative
renormalization group arguments, we expect that the nonconstancy
of the coupling produces at worst a factor of the form
$\exp[C'(m)(\ln r_{ij})^{\alpha(m)}]$ on the right hand side,
where $\alpha(m)<1$ is an $m$-dependent exponent. If the twist
operator does not mix with any other operator in the RG, then the
factor is only an $m$-dependent power of $\ln r_{ij}$.

In general at a random critical point, the logarithm of any
correlation function is expected to have mean and variance
depending logarithmically on the distance; the coefficients of
these logarithmic dependences are universal. This arises because
each extra factor of (say) 2 in distance is expected to contribute
identically-distributed, essentially independent factors to the
correlation function. The central limit theorem then applies to
the distribution as $r_{ij}\to \infty$. (Here we assumed the
moments exist. If the distribution of the logarithm of the factors
in the correlation function is too broadly distributed for this to
hold, then there is still a limiting distribution with universal
properties, in particular the mean (or center) of the distribution
varies as $\ln r_{ij}$, and the width increases as a universal
power, between $1/2$ and $1$, of $\ln r_{ij}$, both with universal
coefficients. For a more general discussion of the scaling forms,
not assuming the product ansatz, see Ref.\ \cite{ludwig90}.) In
our weakly-coupled Gaussian field theory, the moments in Eq.\
(\ref{mom}) have the form we would obtain by assuming the log of
the squared correlation function is Gaussian-distributed; the mean
and variance we would obtain are
\begin{eqnarray} \overline{\ln(\langle \sigma_i\sigma_j\rangle)^2}&=& {\cal
O}([\ln r_{ij}]^{\alpha'}), \nonumber\\ \overline{[\ln(\langle
\sigma_i\sigma_j\rangle)^2]^2}&=& \ln r_{ij}+ {\cal O}([\ln
r_{ij}]^{\alpha''}), \label{momlog} \end{eqnarray} where
$\alpha'$, $\alpha''$ (both $<1$) are again some exponents. Thus
these resemble the results for a random critical point, if we
ignore the possible subleading corrections. Although it is
well-known that the log-normal distribution is not uniquely
defined by its moments, it is plausible that in the present
problem the distribution is indeed asymptotically log-normal. Some
consideration of diagrams for directly disorder-averaging powers
of the logarithm of the ratio of determinants in some models,
using the self-consistent Born approximation to obtain
weak-coupling, also suggests that this is correct (the normal
distribution is uniquely defined by its moments). Note that
strictly we considered the limit $g^2\to 0$ (or $r_{ij}\to\infty$)
for each $m$; this suffices to obtain ``weak convergence'' of the
distribution. At fixed $g^2$ or $r_{ij}$, high moments, or the
tails of the distribution, may not conform to the (log-) normal
form.

The fact that Eq.\ (\ref{mom}) eventually exceeds $1$ implies that
{\em this behavior is impossible in any RBIM with positive
Boltzmann-Gibbs weights}. The metallic phase in class D/B/BD
cannot occur in such a model. Instead, there can presumably be
only gapped or localized phases and critical points between them
(and possibly critical phases, meaning regions with scale
invariance but described by a non-trivial fixed-point field
theory, not a weakly-coupled nonlinear sigma model)---unless some
other, so far unknown, stable metallic phase with the symmetries
of the RBIM exists, that avoids the contradiction found here. This
applies to the zero, as well as the nonzero, temperature region.
As we saw, the regulated spin correlation goes to one as $T\to 0$,
for any distribution of bonds. Even though this is not the same as
the correct, $\eta=0$, correlation for certain bond distributions,
it is still in disagreement with the metallic phase.

We emphasize that results of a similar form can be obtained for
the moments of the twist correlations in a variety of other
metallic regimes in different ensembles, since these are by
definition regions of diffusive behavior that can be described by
a nonlinear sigma model at weak coupling. This is true even in
systems that do not renormalize towards weak coupling, on length
scales shorter than that for the crossover to strong coupling. Two
other cases, class DIII and the symplectic (e.g.\ spin-orbit
scattering) case of the Wigner-Dyson ensembles, both of which
possess Kramers degeneracy due to time-reversal symmetry, flow to
weak coupling in 2D like the class D/B/BD case considered here.
However the physical significance of the Ising order correlation
is less clear in these systems. Another case of interest is a
family of nonlinear sigma models with target space
SO($2n+1$)/U($n$), which with $n\to 0$ arose in connection with
the Nishimori line \cite{grl}. The $m=$ even moments of $\langle
\sigma_i \sigma_j\rangle$ can be considered in this case also. The
twist operator has the same form, but the total number of
Goldstone modes is different: because SO($2n+1$)/U($n$) is the
same, as a manifold, as SO($2n+2$)/U($n+1$), it is as above but
with $n\to1$, not $0$. The family of models with SO($2n+1$)
symmetry has two coupling constants in place of $g^2$ \cite{grl},
but these do not enter the twist correlation at the Gaussian
level. Thus the above result (\ref{mom}) can be used with $n\to
1$, and the moments go as $\sim r_{ij}^{m(m-2)/8}$. For $m>2$,
these increase with $r_{ij}$, eventually exceeding 1, requiring
that $\langle \sigma_i \sigma_j\rangle^2>1$ with nonzero
probability. Thus the weak-coupling region of this family of sigma
models is inaccessible in a RBIM with real couplings. There might
in principle be metallic regimes in other weakly-coupled nonlinear
sigma models in which the original Ising correlations are
represented by a different sort of twist operator that gives a
different result, but we are unaware of any at present.

According to recent work, certain network models are believed to
possess a metallic phase \cite{ohone}. The model of Cho and Fisher
\cite{cf} is equivalent to an Ising model with couplings $\pm K$
on horizontal links (see Fig.\ \ref{fig1}), and $K$, $K+i\pi/2$ on
vertical links \cite{grl}, with independent probabilities $1-p$,
$p$ ($K$ is positive, and $p$ was denoted $W$ in Ref.\ \cite{cf}).
{}From our remarks in Sec.\ \ref{ferm}, this can also be rephrased
by saying that in the Cho-Fisher model, $\pi$ fluxes are added
randomly in pairs, one above the other in Fig.\ \ref{fig1}, on
{\em both} sublattices of plaquettes \cite{cf}. On the $p=1/2$
line, the Cho-Fisher model is equivalent by a gauge transformation
to the O(1) model described in Sec.\ \ref{twist} above
\cite{ohone}. [The equivalence holds in the bulk but breaks down
when we consider the boundary conditions; for certain boundary
conditions, the Cho-Fisher model has the symmetries strictly of
class D ($1-{\cal U}$ has no exact zero eigenvalues).] In Ref.\
\cite{ohone}, the Cho-Fisher model was re-examined numerically,
and metallic behavior was found in a region including the whole of
the $p=1/2$ line. We expect therefore that that model flows to the
weak-coupling regime of the class D/B/BD sigma model, and then the
above result applies. Hence we see that this does not contradict
our claim that no metallic phase can occur in RBIMs with positive
Boltzmann-Gibbs weights. The result for the O(1) model is not
really surprising, in view of the behavior seen above in the dual
of the RBIM, in which the Kramers-Wannier spins $\mu_\alpha$ have
couplings $\tilde{K}_{\alpha\beta}$ with imaginary parts, and the
moments of their correlations can be larger than one.

\section{Correlations in the O(1) Model}
\label{O(1)}

Our final result is for the O(1) model, already introduced in
Sec.\ \ref{twist}. We will argue that it is never in the ordered
or disordered phases of the Ising model, by showing that the
moments of the squared order and disorder correlations are both
bounded below by 1. We also point out that the latter behavior is
found in the other network models, in classes A, C. In the class C
(spin quantum Hall) case, we find the exact exponent for the mean
order and disorder correlations at criticality.

In the O(1) model, or the Cho-Fisher model at $p=1/2$, each
plaquette of the network model (medial graph of the Ising model
square lattice) encloses a flux of either $0$ or $\pi$ with
independent probabilities $1/2$ (up to some boundary effects). We
consider the order or disorder correlation functions, defined as
before in fermion language. Like the disorder correlations in the
RBIM with symmetric distribution of $J_{ij}$'s, {\em the logarithm
of either squared correlation is symmetrically distributed, and
the even moments of the correlations are bounded below by 1}. Note
that this behavior is consistent with our results in Eq.\
(\ref{momlog}), if the error term in
$\overline{\ln(\langle\sigma_i\sigma_j\rangle)^2}$ is zero. This
means that if the O(1) model really does flow to the metallic
phase, then these universal correction terms, and all higher-order
analogs, in the nonlinear sigma model, must be zero.

Now we compare this with the behavior expected in the localized
phases. Such phases occur at weak disorder (small $p$) in the RBIM
and Cho-Fisher models. Like the two phases of the pure Ising
(massive Majorana field theory) model, one or other mean (and mean
square) correlation is supposed to decay to zero, and the other to
go to a constant, at large distance. Hence the O(1) model is
definitely not in either such phase. It is tempting to conclude
that it must therefore be in the metallic phase, though this is
not really proved; some other localized phase may not be ruled
out. As mentioned above, numerical work \cite{ohone} led to the
hypothesis of metallic behavior everywhere in the O(1) model.

The reason for caution about the last point is obtained by
considering other network models for other ensembles. The two
models in question are defined similarly to the O(1) model, but in
the first, the particles pick up independent,
uniformly-distributed U($1$) phases on the links, and in the
second they pick up SU($2$) matrices instead (the latter requires
two-component wavefunctions for the particles). These are
respectively the Chalker-Coddington model for the integer quantum
Hall transition (class A) \cite{cc}, and the Kagalovsky {\it et
al.} model for the spin quantum Hall transition (class C)
\cite{kag}. Both models possess localized phases away from their
critical points. We now consider twist (``order'' or ``disorder'')
correlations, defined as the ratio of modified to unmodified
partition functions (fermion determinants) as before. First we
note that for the class A model, $\cal U$ is a $4N\times 4N$
unitary matrix, its eigenvalues come in pairs $e^{-i\epsilon}$,
$-e^{-i\epsilon}$, and $\det(1-{\cal U})$ is in general complex.
For class C, $\cal U$ is an $8N\times 8N$ symplectic matrix, and
its eigenvalues come in quadruplets, $e^{-i\epsilon}$,
$-e^{-i\epsilon}$, $e^{i\epsilon}$, $-e^{i\epsilon}$, similar to
class D; hence $\det(1-{\cal U})$ is real and positive. This
applies to both the modified and unmodified partition functions,
and we see that in the U($1$) (class A) case we should consider
the modulus square correlations, while for the SU($2$) (class C)
case we can consider the correlations themselves, which are real
and positive. The uniform distributions imply independent uniform
distributions for the flux (in U(1) or SU(2) respectively) through
each plaquette in these models. Multiplying these fluxes (as group
elements) by $-1$ (for the twist insertion operation) leaves the
distributions unchanged, and hence again the logarithm of the
[modulus squared in the U($1$) case] order or disorder
correlations in these models are symmetrically-distributed, and
the moments of the [mod-squared for U($1$)] correlations are
bounded below by 1, {\em even in the localized phases}.

Thus it appears that these localized phases behave differently
from those in the RBIM and Cho-Fisher models, and cannot be
distinguished by Ising order or disorder variables. This may be
connected with the continuous distributions for the flux in the
plaquettes in these models, as opposed to the discrete
distributions for the flux (which was either 0 or $\pi$) in the
O(1) model. However, we may also point out that the localized
phases in the RBIM and Cho-Fisher models, like the localized
phases of a Majorana Fermi field with a weakly random mass, are
expected to have vanishing density of states at $\epsilon=0$, at
least at weak disorder. In contrast, there is a possibility of a
localized phase with the statistics of class D/B/BD, and the mean
{\em local} density of states (which is independent of system
size) near $\epsilon=0$ would be expected to be nonvanishing [as
in the localized phase in the U(1) network model, which is in the
unitary (class A) ensemble, though not the SU(2) model which is in
class C] and smoothly varying. Further, it would probably have an
$\epsilon$ dependence like that for class D in Ref.\ \cite{bsz},
including a peak at $\epsilon=0$, but with the energy scale for
such structure proportional to the inverse-square localization
length. Although the existence of such a localized phase was
predicted in, for example, Ref.\ \cite{readgr}, it has not so far
been demonstrated to occur in practice in any model. In fact,
given the symmetries of the problem, it is not clear why such
behavior does not occur in the localized phases of the RBIM, or
for the Majorana fermion with random mass. Perhaps such a phase
would be consistent with the above form of probability
distribution, and have neither order nor disorder correlations
decaying to zero. If so, it may, like the metallic phase, be
inconsistent with the Ising correlations in a RBIM.

Since we have been discussing the class C (spin quantum Hall)
network model \cite{kag}, we also include here a result for the
critical correlations of the Ising order and disorder operators in
that model. We can obtain a result only for the mean values,
$\overline{\langle\sigma_i\sigma_j\rangle}$ and
$\overline{\langle\mu_\alpha\mu_\beta\rangle}$. Each of these is
defined by a twist of each of the two components of the
wavefunction in a single copy of the system. It will be necessary
here to use the supersymmetry method, in which the division by the
unmodified partition function for a single copy is represented by
a single two-component boson field \cite{glr,sf1}. The partition
function of the unmodified supersymmetrized system has
supersymmetry osp($2|2$) $\cong$ sl($2|1$) \cite{glr}, and is
equal to 1. The mean of either correlation is represented by a
modified partition function, in which two twist operators have
been inserted. The partition function, like the unmodified one,
has a graphical expansion as a sum over coverings by
nonintersecting loops on the network model (medial) graph, with
certain weights. States of the fermions and bosons flow around the
loops; there are only three possible states, which can be labeled
by the number of fermions they contain, either $0$, $1$, or $2$
\cite{glr,sf1}. In the unmodified partition function, each loop is
weighted with a factor 1, because the singly occupied state
contributes $-1$, and the other two states $+1$ each. (In Ref.\
\cite{glr}, this mapping was constructed and used to show that
several exponents for the spin quantum Hall transition are given
by percolation, which has the identical loop expansion.) Because
the original twist weights a fermion of either component that
propagates once around the twist with a factor of $-1$, the
singly-occupied state picks up a $-1$ and the others are
unchanged. That is, a loop that encircles one of the twist
insertions and not the other is now weighted by a factor $3$, not
1; the other factors which occur at the nodes \cite{glr} are
unchanged. It follows that either mean correlation is greater than
one, as we proved by another method already. Since the maximum
number, and the typical number, of such loops will increase with
the separation, without limit in the critical case, we expect that
either correlation increases as $\sim r^{-2x}$, where $x<0$ is the
scaling dimension of the twist operator. In the loop model
description, the twist operator has exactly the form recently
considered by Cardy \cite{cardy} for percolation and other
problems. Making use of his Eq.\ (1), $x=(\chi'^2-\chi^2)/(2g)$,
with $g=2/3$, $\chi=1/3$ for percolation, and $2\cos \pi\chi'=3$
for our twist, we obtain
\begin{equation}
x=-\frac{1}{12}-\frac{3\ln^2[(3+\sqrt{5})/2]}{4\pi^2}\simeq
-0.154. \end{equation} It is implicit in this result that we chose
the branch for the logarithm such that $x$ is real. With this
choice, Cardy's general formulas imply that $x$ is negative
whenever the factor for loops that enclose exactly one twist
operator is larger than that, $2\cos\pi\chi$, for the unmodified
loops.

\section{Applications to superconductors and paired FQHE states}
\label{app}

Paired states of fermions with complex (time-reversal violating)
pairing of spinless or spin-polarized particles, or systems with
broken time-reversal symmetry and spin-orbit scattering, have the
same symmetries as class D/B/BD \cite{sf,readgr}. We will consider
only one-component systems such as spinless or spin-polarized
fermions with p-wave pairing, which are the simplest, and begin
with the pure case. We will then argue that their phase diagrams
are more like that of the RBIM than has previously been
recognized. For a RBIM in which frustrated plaquettes (of the
Ising model lattice) are introduced independently, with some
density, we argue that the Ising ordered phase is destroyed at
$T>0$ for an arbitrarily small density of frustrated plaquettes
(vortices). In the FQHE, this implies that the weak-pairing
(nonabelian statistics) phase is destroyed by weak disorder.

It was important in Ref.\ \cite{readgr} for the discussion of
nonabelian statistics that vortices carry a Majorana fermion zero
mode when they occur in the so-called weak-pairing phase, but not
in the strong-pairing phase. These phases occur on the two sides
of the transition at which the mass of the Majorana fermions
changes sign; the weak-pairing phase corresponds to the Ising
ordered (low $T$) phase. The notion of a dual (in the Ising sense)
vortex with the opposite properties---i.e.\ carrying a fermion
zero mode only in the strong pairing phase---was implicitly
discarded. The two types of vortices correspond in the network
model to the two sublattices of plaquettes on which vortices (or
fluxes) may be added to those already present in the pure model.
The first type corresponds to adding a vortex on the network model
plaquettes that correspond to plaquettes of the Ising model. In a
continuum model, only the first type of vortex was considered
because it was argued that in the physical situation a vortex
should effectively have a region of strong-pairing phase, or
vacuum (which was argued to be effectively the same thing in a
``topological'' sense), at its core.

The argument can be made essentially rigorous by considering a
tight-binding Bogoliubov Hamiltonian on a lattice with a finite
number $N$ of sites, with an edge, not periodic boundary
conditions. Such a Hamiltonian corresponds to a $2N\times 2N$
matrix, which is in the Lie algebra of SO($2N$), and its
eigenvalues come in pairs $\epsilon$, $-\epsilon$, so that it
never has an odd number of exact zero eigenvalues. A vortex can be
generally defined as an object which, far from its core,
approaches a singular gauge transformation that describes the
insertion of a flux $\pi$ into the system without the vortex; this
means that both the phase of the gap function, and the vector
potential exhibit the winding by $\pi$. If we insert one of the
postulated dual vortices in the strong-pairing phase with no other
vortices present, then it is supposed to carry a zero-energy mode.
There are no other zero modes with which it can mix, as we know
because far from the vortex, we can use our understanding of the
low-energy properties. In particular, there is no chiral spectrum
of edge excitations in the strong-pairing phase. Hence it must be
an exact zero mode, which is impossible for this Hamiltonian. A
similar argument can be given in the weak-pairing phase, where the
dual vortex does not carry a zero mode, but induces one on the
edge that encloses it. We conclude then that no such dual vortices
can exist, and there is only one type of vortex. Clearly we may
take a continuum limit and draw the same conclusion. This means
that Kramers-Wannier duality does not apply in such Hamiltonian
models.

Turning to quenched disorder, it was shown in Ref.\ \cite{readgr}
that randomly-inserted vortices in independent positions are a
relevant perturbation of the pure Ising (Majorana) fixed point
theory \cite{ye}. It was pointed out that such disorder always
occurs in the applications to FQHE systems, where underlying
potential disorder can induce vortices, because they are charged.
(In the RBIM, it is well-known that the vortices are correlated in
pairs.) It seemed natural to expect such disorder to cause a flow
to the metallic phase in class D/B/BD. Now if we assume that all
the vortices are of the first type defined above, then we can
construct network models of this situation by adding $\pi$ fluxes
independently, all on the same sublattice, with some density $p$.
This can be described as an Ising model with bond-disorder that is
not independent for each bond, but the bonds are still real, and
of fixed magnitude. This will accurately model the low-energy
properties if the vortices are dilute (note that we assume the
penetration depth and coherence length of the pure system stay
finite at the transition). The results of this paper show that
{\em such a model cannot have the metallic phase in class D/B/BD}.

In fact, such a model may not have a phase transition either.
Introducing frustrated plaquettes into the Ising model
independently tends to destroy Ising long-range order, though the
discussion is complicated by the gauge choices needed in placing
the strings of negative bonds that are needed to produce the
frustrated plaquettes. We can avoid this difficulty by considering
the spin-glass order parameter or correlations instead. Because
the bonds are $\pm J$, there will be ground state degeneracy.
Ground states can be represented by lines of frustrated bonds that
join the frustrated plaquettes in pairs, chosen so as to minimise
the energy. Distinct ways of dividing the frustrated plaquettes
into pairs will frequently be exactly degenerate, and reconnecting
the lines of frustrated bonds means reversing the spins in some
domain. A condition that plausibly is necessary, but may not be
sufficient, for the absence of long-range order at $T=0$, and
hence of a finite $T$ phase transition, is that, in a ground state
in the thermodynamic limit, any given spin lies, with probability
1, in a finite domain that can be flipped with zero energy cost.
Heuristic considerations of sufficiently large domains suggest
that any spin does lie in at least one such flippable domain (the
probability for a domain, formed by reconnections of lines of
frustrated bonds between nearby frustrated plaquettes, to have
zero energy cost decreases as a power of the length of its
boundary, while the number of such domains is exponential in this
length). Hence we suspect that there is no ordering even at zero
temperature in this model, for any nonzero density of the
frustrated plaquettes (however, the zero-temperature state may be
critical, as that of the usual $\pm J$ EA model may be also). This
means that the model at $T>0$ is presumably in the paramagnetic
phase, and all fermion eigenstates are presumably localized. The
model can be generalized by introducing a continuous distribution
for the magnitudes of the $J_{ij}$, and will then order at $T=0$,
but a similar argument for the free energy at finite $T$ suggests
that it will still not order at $T\neq0$.

A stronger argument can also be given. The mean ground state
energy is a function of the density $p$ of frustrated plaquettes,
and is extensive, and varies smoothly with $p$ except at $p=0$.
Increasing $p$ slightly means frustrating a small number of
previously unfrustrated plaquettes. Thus frustrating one
additional plaquette changes the mean total energy by an amount of
order one. Yet this forces a domain wall from the plaquette to the
edge of the system, along which the lines of frustrated bonds have
been reconnected. The only reasonable conclusion is that the mean
energy of the minimum energy domain wall is zero, except for a
finite effect from around the added frustrated plaquette. This is
the same behavior as in an EA spin-glass model. A similar
conclusion holds for two added frustrated plaquettes, and may be
compared with the discussion in Sec.\ \ref{symm} (but note that
there the operation also unfrustrated originally-frustrated
plaquettes, so that for a symmetric distribution of bonds the mean
free energy change was exactly zero). We have no information about
the width of the distribution of the domain wall energies, but we
can expect that, like the usual short-range EA 2D spin glass
models, there will be no finite $T$ transition. (A finite $T$
transition can occur in a 2D spin glass with sufficiently
long-range power-law random bonds, but we have no reason to expect
this to be realized here.)

Intuitively, the independently-inserted vortices appear similar to
a random magnetic field, though the relation is not exact.
However, it is a field that couples to the dual variables
$\mu_\alpha$, and further it has a uniform component. The latter
is the most important effect. A uniform magnetic field in a
ferromagnetic Ising model destroys the transition, and the
resulting phase has correlations like those in the ordered phase
for the spin to which the field couples. In the present case, it
would lead to the high temperature paramagnetic phase, in
agreement with our conclusion.

The conjecture that vortex disorder destroys the Ising ordered
phase has a dramatic consequence for applications to spinless or
spin-polarized FQHE paired states. The paramagnetic phase
corresponds to the disordered version of the strong-pairing phase;
the weak-pairing phase has been destroyed. The weak-pairing or
Moore-Read phase was supposed to be the basis for nonabelian
statistics \cite{readgr}. We are arguing that this behavior,
including the chiral Majorana fermion edge modes, is destroyed by
any weak vortex (i.e.\ potential) disorder.

In models, such as the tight-binding Bogoliubov Hamiltonian, in
which the vector potential and gap function each has only short
range correlations, vortices will tend to be produced only in
pairs, and again can be of only one type. Thus these one-component
models appear similar to the RBIM, and may have a similar phase
diagram, in which the weak-pairing phase occurs at weak but not at
strong disorder. At the transition, the critical behavior may be
that of either the pure Ising model (up to logarithms) or the low
$T$ phase boundary in the (frustrated) RBIM. Thus the latter
universality class could be realized in one-component
superconductors. We conclude (in contrast to Refs.\
\cite{sf,readgr}) that, in at least some models of superconductors
or FQHE paired states with the symmetries of class D/B/BD, and
with only one type of vortex present, there may not be a metallic
phase after all, but there may instead be a transition in a
distinct universality class from the pure system for some types of
disorder.

Clearly, similar possibilities should be explored in connection
with other ensembles, which can occur when more components are
present, but will not be considered further here. We point out,
however, that the case of pairing of spin-1/2 fermions, with
spin-orbit scattering and a general random mass, which has the
symmetries of class D/B/BD, corresponds to multicomponent models
considered in Ref.\ \cite{bsz}, where it is argued that a metallic
phase is produced. Ref.\ \cite{bsz} also argued that the metallic
phase would not occur in the one-component case in the absence of
vortex disorder, but did not consider vortex disorder as fully as
we have. Also, models of superconductors as disordered grains
(each described by a random matrix from class D), coupled by weak
hopping, appear similar to multicomponent models, and may have a
metallic phase, as a mapping to a weakly-coupled nonlinear sigma
model would suggest.

\section{Conclusion}
\label{con}

The main point to emerge from this study is that the important
difference between the RBIM and the (one-component) models which
possess a metallic phase is that in the former the added vortices,
in network model language, occur on one sublattice only, but on
both in the latter. This appears to be a necessary condition for
the existence of the metallic phase. If the vortices are
correlated in pairs, sufficiently strong disorder will be required
to produce the metallic phase. This result casts some doubt on
whether the metallic phase will occur in applications to
one-component 2D superconductors and paired FQHE systems, because
these possess only one type of vortex, corresponding to those on
only one sublattice in the network. On the other hand, if vortices
of only one type are present, but are uncorrelated, this may lead
to the destruction of one of the phases, and hence of the phase
transition. If such vortices are correlated in pairs, the phase
diagram may resemble that of the RBIM. It would be interesting to
test this numerically, both on models defined by a Hamiltonian,
such as tight-binding models, and on network models, and also for
other symmetry classes.

\acknowledgments

We are grateful to I. A. Gruzberg, J. T. Chalker, M. R. Zirnbauer,
and D. S. Fisher for discussions. N.R. is grateful to David Gross
and the staff of the Institute for Theoretical Physics, University
of California, Santa Barbara, for their hospitality while this
paper was being written. This work was supported by the NSF under
grant no.\ DMR-98-18259 (NR). Work of N.R. at the ITP was
partially supported by the NSF, under grant no.\ PHY94-07194.

\end{document}